\documentclass[twocolumn,showpacs,preprintnumbers,amsmath,amssymb]{revtex4}

\usepackage{graphicx}
\usepackage{rotating}
\usepackage{dcolumn}% Align table columns on decimal point
\usepackage{bm}% bold math

\begin{document}

\draft

\title{Dynamics of a strongly interacting Fermi gas: the radial quadrupole mode}

\author{A. Altmeyer,$^{1}$ S. Riedl,$^{1,2}$ M. J. Wright,$^{1}$ C. Kohstall,$^{1}$ J. Hecker Denschlag,$^{1}$ and R. Grimm$^{1,2}$}
\address{$^{1}$Inst.\ of Experimental Physics and Center for Quantum Physics, Univ.\ Innsbruck,
%Technikerstra{\ss}e 25,
6020 Innsbruck, Austria\\$^{2}$Inst.\ for Quantum Optics and Quantum
Information, Acad.\ of Sciences, 6020 Innsbruck, Austria}
\date{\today}

\begin{abstract}

We report on measurements of an elementary surface mode in an
ultracold, strongly interacting Fermi gas of $^{6}$Li atoms.  The
radial quadrupole mode allows us to probe hydrodynamic behavior in
the BEC-BCS crossover without being influenced by changes in the
equation of state. We examine frequency and damping of this mode,
along with its expansion dynamics. In the unitarity limit and on the
BEC side of the resonance, the observed frequencies agree with
standard hydrodynamic theory. However, on the BCS side of the
crossover, a striking down shift of the oscillation frequency is
observed in the hydrodynamic regime as a precursor to an abrupt
transition to collisionless behavior; this indicates coupling of the
oscillation to fermionic pairs.

\end{abstract}



\maketitle
%******************************************************************************************
\section{Introduction}
\label{introduction}
%******************************************************************************************
The advent of ultracold, strongly interacting Fermi gases
\cite{Ohara2002,Bourdel2003}, molecular Bose-Einstein condensates
\cite{Jochim2003c,Greiner2003,Zwierlein2003}, and fermionic
condensates \cite{Regal2004,Zwierlein2004} has opened up unique
possibilities to study the fundamental physics of interacting
fermions. The availability of controllable model systems with
tunable interactions provides unprecedented experimental access to
the many-body physics of fermionic quantum systems, which is of
great fundamental importance for various branches of physics
\cite{Varenna2006}.

A fundamental problem, which has been discussed in the theoretical
literature for decades
\cite{Eagles1969,Leggett1980,Nozieres1985,Engelbrecht1997}, is the
crossover from Bose-Einstein condensation (BEC) to a macroscopic
quantum state in the Bardeen-Cooper-Schrieffer (BCS) regime. In this
crossover, the nature of pairing changes from the formation of
bosonic molecules by fermionic atoms to pairing supported by
many-body effects. With novel model systems now available in
ultracold Fermi gases, the BEC-BCS crossover has recently stimulated
a great deal of interest in both theory and experiment
\cite{Varenna2006}.

Collective excitation modes in trapped ultracold Fermi gases provide
powerful tools to investigate the macroscopic properties of a system
in the BEC-BCS crossover \cite{Varenna_Rudi2006}. For experiments of
this class, ultracold $^6$Li gases have excellent properties. This
is because of their stability in the molecular regime
\cite{Cubizolles2003,Jochim2003a,Jochim2003c} and precise magnetic
tunability of interactions based on a broad Feshbach resonance
\cite{Houbiers1998a,Bartenstein2005}. Early experiments on
collective modes in the BEC-BCS crossover provided evidence for
superfluidity \cite{Kinast2004} and showed a striking transition
from hydrodynamic to collisionless behavior \cite{Bartenstein2004b}.
More recent experiments yielded a precision test of the equation of
state \cite{Altmeyer2006a}. The previous experiments have focussed
on collective modes with compression character, where both the
hydrodynamic properties and the equation of state determine the mode
frequency
\cite{Kinast2004,Bartenstein2004b,Kinast2004a,Kinast2005,Altmeyer2006b,Altmeyer2006a}.

In this Article, we report on measurements of a {\it pure surface
mode} in the BEC-BCS crossover, which provides new insight into the
dynamics of the system. The ``radial quadrupole mode'' in an
elongated trap, the fundamentals of which are discussed in
Sec.~\ref{theory}, allows for a test of hydrodynamic behavior
without being influenced by changes in the equation of state. In
Sec.~\ref{experimental procedure}, we present our experimental setup
and the main procedures. We introduce a tool to excite collective
oscillations with an acousto-optic scanning system. The results of
our measurements, presented in Sec.~\ref{results}, provide us with
new insight on the abrupt transition from hydrodynamic to
collisionless behavior, first observed in \cite{Bartenstein2004b}.
The present work provides strong evidence that quasi-static
hydrodynamic theory \cite{Varenna_Stringari2006} does not apply to
collective modes of a strongly interacting fermionic superfluid,
when the oscillation frequencies approach the pairing gap
\cite{Chin2004a}.
%\newpage
%**********************************************************************************
\section{Radial Quadrupole Mode}
\label{theory}
%**********************************************************************************
The confining potential in our experiments is close to the limit of
an elongated harmonic trap with cylindrical symmetry. In this case,
we can consider purely radial collective oscillations, neglecting
the axial motion. The frequencies of the radial modes can be
expressed in units of the radial trap frequency $\omega_{\rm r}$. We
note that our experiments are performed in a three-dimensional
regime, where the energy $\hbar \omega_{\rm r}$ %($\hbar$ is Planck's
%constant divided by $2\pi$)
is typically a factor of 30 below the
chemical potential and finite-size effects can be neglected.

\begin{figure}[h]
\center{
\includegraphics[width=8.5cm]{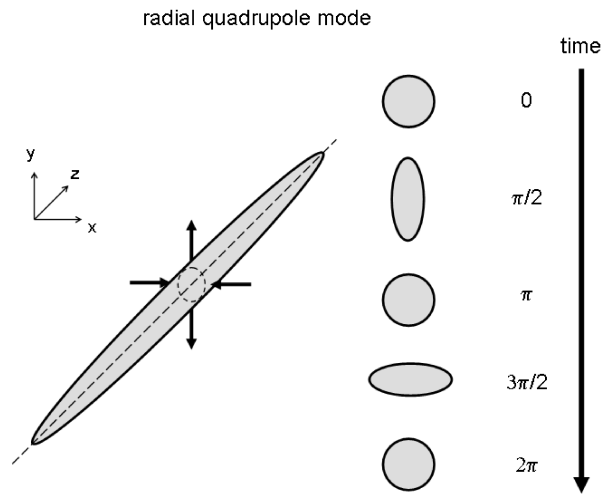}}
\caption{Illustration of the radial quadrupole mode as an elementary
collective excitation of an elongated, trapped atom cloud.}
\label{surface_mode_overview}
\end{figure}

In this situation, there are two elementary collective modes of the
system, the radial compression mode and the radial quadrupole mode
\cite{Stringari1996a,Varenna_Rudi2006}. We focus on the quadrupole
mode, which is illustrated in Fig.~\ref{surface_mode_overview}. This
mode corresponds to an oscillating radial deformation, which can be
interpreted as a standing surface wave. The mode was first
demonstrated in atomic BEC experiments \cite{Onofrio2000a} and
applied to investigate rotating systems \cite{Bretin2003a}, but so
far it has not been
studied in strongly interacting Fermi gases.\\
Being a pure surface mode, the frequency $\omega_{\rm q}$ of the
radial quadrupole mode does not depend on the compressibility of the
system. The frequency $\omega_{\rm q}$ does not depend on the
equation of state but on the collisional properties. In the
hydrodynamic regime, whether the gas is a superfluid or a classical
gas with a collision rate strongly exceeding the radial trap
frequency, the frequency of this mode is given by
\cite{Stringari1996a}
\begin{equation}
\label{paper_surface_freq_hydrodynamic} \omega_{\rm q} =
\sqrt{2}\,\omega_{\mathrm{r}}.
\end{equation}
In contrast, for a collisionless gas, where the atoms freely
oscillate in the trap, the frequency is
\begin{equation}
\label{paper_surface_freq_collisionless} \omega_{\rm q} =
2\,\omega_{\mathrm{r}}.
\end{equation}

Because $\omega_{\rm q}$ is insensitive to the compressibility of
the gas and the difference between the collisonless and the
hydrodynamic frequencies is large, the radial quadrupole mode can
serve as an excellent tool to probe pure hydrodynamics. Particularly
interesting is the transition from hydrodynamic to collisionless
behavior at lowest temperatures. Such a change occurs in a strongly
interacting Fermi gas on the BCS side of the resonance
\cite{Bartenstein2004b,Kinast2004a,Altmeyer2006b}. Near this
transition, measurements on the compression mode indicated frequency
down shifts, which raised questions concerning the validity of
standard hydrodynamic theory in this interaction regime
\cite{Combescot2004a,Combescot2006}. Previous experiments could not
unambiguously identify the origin of frequency shifts near the
hydrodynamic-to-collisionless transition, which is a particular
motivation for probing the crossover gas with the radial quadrupole
mode.

%\newpage
%***********************************************************************
\section{Experimental Procedure}
\label{experimental procedure}
%***********************************************************************
The apparatus and the basic preparation methods for experiments with
a strongly interacting Fermi gas of $^{6}$Li atoms have been
described in our previous work
\cite{Jochim2003c,Bartenstein2004a,Bartenstein2004b,Chin2004a}. As a
starting point, we produce a molecular BEC of $^{6}$Li$_{2}$
\cite{Jochim2003c,Bartenstein2004a}. By changing an external
magnetic field, we can control the inter-particle interactions in
the vicinity of a Feshbach resonance, which is centered at $834$G
\cite{Houbiers1998a,Bartenstein2005}. The interactions are
characterized by the atomic s-wave scattering length $a$.

We start our experiments with an ensemble of about $N =
4\times10^{5}$ atoms in an almost pure BEC at a magnetic field of
$764~$G. In order to change the properties of the system
adiabatically, we slowly ramp to the final magnetic field, where the
measurements are performed \cite{Bartenstein2004a}. The temperature
of the gas is typically below 0.1 $T_{\rm F}$, unless stated
otherwise.

In order to observe the collective oscillations, we take absorption
images of the cloud in the x-y-plane after release from the trap. We
illuminate the atoms with a probe beam along the z-direction of the
cigar-shaped cloud. The probe light causes a resonant excitation of
the D2-line, at a wavelength of 671nm. We use dichroic mirrors for
combining and separating the probe and the dipole trapping beam. The
frequency of the probe beam can be tuned over a range of more than
1GHz, which enables resonant imaging over the whole range of
magnetic fields that we create in our experiments.

The gas is confined in a nearly harmonic trapping potential, which
has an axially symmetric, cigar-shaped trap geometry. Optical
confinement in the radial direction is created by a focused 1030-nm
near-infrared laser beam with a waist of $\sim 58~ \mathrm{\mu}$m.
The potential in the axial direction consists of a combination of
optical and magnetic confinement \cite{Jochim2003c}; the magnetic
confinement is dominant under the conditions of the present
experiments. We set the laser power to 270 mW, which results in a
radial trap frequency of $\omega_{\mathrm{r}} \approx 2\pi \times
370~$Hz and an axial trap frequency of $\omega_{\rm z} \approx 2\pi
\times
 22~$Hz at a magnetic field of
$764~$G. The trap frequencies correspond to a Fermi energy of a
noninteracting cloud $E_{\rm F} = \hbar (\omega_{\rm
r}^{2}\omega_{\rm z}3N)^{1/3} = k_{\rm B}\times
 740$nK.

In order to excite collective oscillations, we suddenly change the
optical trapping potential. The position and shape of our trapping
potential in the x-y-plane can be manipulated through the use of a
two-dimensional scanning system. One feature of the system is that
we can rapidly displace the trap laterally. Fast modulation of the
beam position enables us to create time-averaged potentials
\cite{Milner2001,Friedman2001}.\\
\begin{figure}[h] \center{
\includegraphics[width=8.5cm]{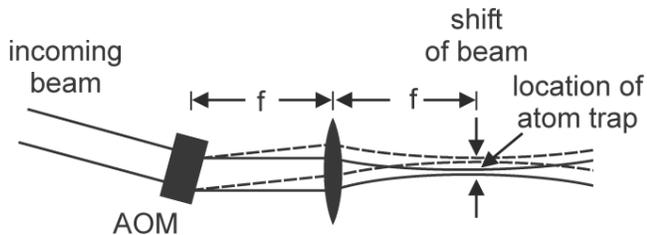}}
\caption{Schematic illustration of the scanning system.  A wide
collimated beam passes through an AOM.  The resulting deflection
angle depends on the driving frequency of the AOM. The beam passes
through a lens at the distance of one focal length behind the AOM.
The lens focuses the beam for atom trapping. A change in deflection
angle results in a parallel shift of the beam position in the focal
plane. The solid and dashed lines show the beam path for different
deflection angles. The zeroth order beam is not shown.}
\label{scanning_system_detail}
\end{figure}
The scanning system is constructed by use of two acousto-optic
modulators (AOMs), which are aligned for vertical and horizontal
deflections. Fig.\,\ref{scanning_system_detail} illustrates the
principle of our scanning system for one direction. A collimated
beam passes through an AOM and is deflected depending on the driving
frequency. A lens is placed at a distance of one focal length behind
the AOM, so that the deflection results in a parallel displacement
of the beam. By changing the driving frequency of the AOM, the
lateral position of the focus is shifted. This system enables us to
displace the focus of the trapping beam in the horizontal and the
vertical direction by up to four times the beam waist in all
directions. Furthermore, the deflection can be modulated by
frequencies of up to $\sim1$MHz within 3dB bandwidth. In our trap
configuration, we use modulation frequencies of $100$kHz, which
greatly exceeds the trap frequency. We create elliptic potentials,
i.e. potentials with $\omega_{\rm x} \neq \omega_{\rm y}$, by
modulating the trap position along a specific direction. We use this
for the excitation of the quadrupole mode. By choosing a suited
modulation function \cite{modulationnote}, these elliptic potentials
are nearly
harmonic.\\
\begin{figure}[h] \center{
\includegraphics[width=9cm]{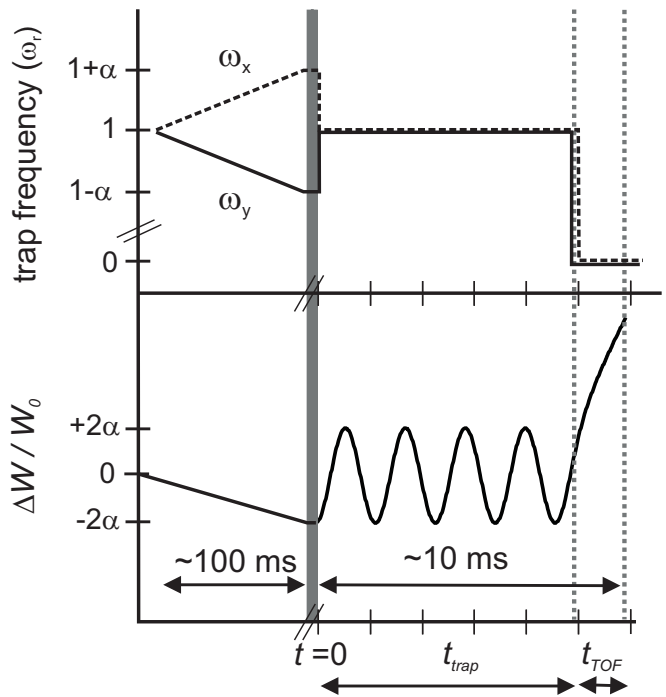}}
\caption{Timing scheme for the excitation of the radial quadrupole
mode. The ellipticity of the trap is slowly ramped up within
100\,ms. This results in a change of $\alpha$ in the trap
frequencies, where $\alpha$ characterizes the ellipticity, and sets
the initial, normalized deformation $\Delta W/W_{0} = -2\alpha$.
$W_{0}$ is defined as the width of the cloud in the trap without
excitation. At $t = 0$, the elliptic deformation is switched off and
the oscillation in the trap begins. (Shown here is an oscillation in
the hydrodynamic regime.) The oscillation continues until the trap
is turned off at $t = t_{\rm trap}$, which is usually between 0 and
10\,ms. At $t=t_{\rm trap}$, the cloud is released from the trap and
expands for the time $t_{\rm TOF}$, which is typically 2\,ms.}
\label{surface_timing}
\end{figure}
When we excite the quadrupole mode, we first adiabatically deform
the trapping potential in $\sim100\,$ms to an elliptic shape. This
slow deformation ensures that the cloud stays in thermal equilibrium
even in the near-collisionless regime and no excitations occur. We
suddenly switch off the deformation leading to an oscillation in the
x-y-plane of the elliptic cloud in the originally round trap.

The initial deformation corresponds to different trap frequencies in
horizontal and vertical direction where $\omega_{ \rm 0x} =
(1+\alpha) \omega_{\mathrm{r}}$ and $\omega_{\rm 0y} = (1-\alpha)
\omega_{\mathrm{r}}$. The parameter $\alpha$ determines the
amplitude of the emerging oscillation; we choose it for most of our
measurements (unless stated otherwise) to be $\alpha \approx 0.05$.
We increase $\alpha$ by increasing the modulation for the time
averaged potential along the y-direction. As the modulation
decreases the confinement strength of the dipole trap, we
 simultaneously ramp up the trap power to ensure that the mean trap
frequency $\omega_{\rm r} = \sqrt{\omega_{ \rm 0x} \omega_{\rm 0y}}$
remains constant. This avoids excitation of the compression mode.

Fig.\,\ref{surface_timing} shows the timing scheme for the
excitation of the radial quadrupole mode. At $t=0$, the collective
oscillation is excited and the cloud starts oscillating in the trap
for a variable time $t_{\rm trap}$. Horizontal and vertical widths
of the cloud, $W_{\rm x}$ and $W_{\rm y}$, oscillate in the trap out
of phase with a relative phase shift of $\pi$. As an observable, we
choose the difference in widths $ \Delta W = W_{\rm x} -
W_{\mathrm{y}}$, which cancels out small effects of residual
compression oscillations. For normalization, we introduce the width
$W_{0}$ of the cloud in the trap without excitation.

Experimentally, we determine the collective quadrupole oscillations
after suddenly switching off the trap and a subsequent expansion
time $t_{\mathrm{TOF}}$. We then take an absorption image of the
cloud and determine its horizontal and vertical widths $W_{\rm x}$
and $W_{\mathrm{y}}$ via a two-dimensional Thomas-Fermi profile fit.
From these measurements after expansion, we can determine the
in-trap behavior.\\
\begin{figure}[h!]
\center{
\includegraphics[width=9.5cm]{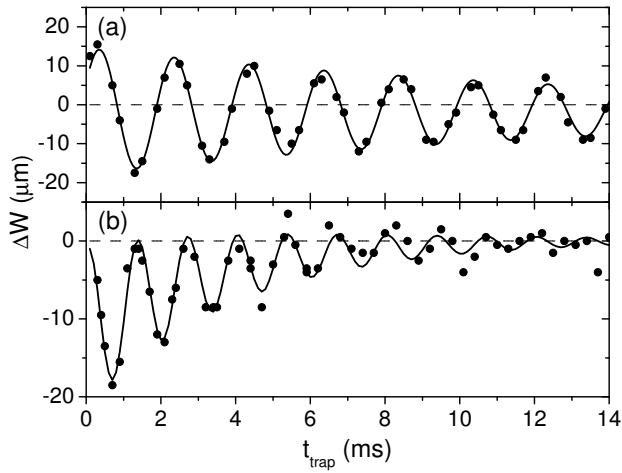}}
\caption{Typical radial quadrupole oscillations in the hydrodynamic
(a) and collisionless (b) regimes.  The solid lines show fits to our
data according to Eq.\eqref{surface_fit}. The dashed lines indicate
$\Delta W = 0$. The expansion time $t_{\mathrm{TOF}}$ is 2 ms. In
(a), the oscillation in the unitarity limit ($B = 834$G) is shown,
whereas (b) shows the oscillations for $B = 1132$G ($1/k_{\rm F}a
\approx -1.34$).} \label{surface_typical_oscillation}
\end{figure}
Typical data sets of radial quadrupole oscillations are shown in
Fig.\,\ref{surface_typical_oscillation}.
Fig.\,\ref{surface_typical_oscillation}(a) shows an oscillation in
the hydrodynamic regime; here we observe a weakly damped harmonic
oscillation centered about a small constant offset.
Fig.\,\ref{surface_typical_oscillation}(b) shows the typical
behavior in the collisionless regime. The frequency of the
oscillation is clearly higher than in the hydrodynamic regime. The
oscillation shows stronger damping and has an exponentially
time-varying offset.

We find that, for both regimes, the dependence of $\Delta W$ on
$t_{\rm trap}$ can be well described by the fit function
\begin{eqnarray}
\label{surface_fit}
\begin{split}
 \Delta W = ~&
A~e^{-\kappa t_{\rm trap}}~\cos{(\omega_{\rm q}t_{\rm trap}
+\phi)}\\& + C~e^{-\xi t_{\rm trap}}+y_{0},
\end{split}
\end{eqnarray}
which is explained in detail in Appendix
\ref{appendix_thermalization}.

Note that the frequency $\omega_{\rm q}$ and the damping constant
$\kappa$ are independent of the expansion during $t_{\mathrm{TOF}}$
and characterize the behavior of the trapped oscillating atom cloud.
In contrast, the amplitude $A$ and the phase shift $\phi$ depend on
the expansion time and provide further information on the dynamics
of the gas. The offset function $C~e^{-\xi t_{\rm trap}}$ with
amplitude $C$ and damping constant $\xi$ results from thermalization
effects and is only relevant in the collisionless regime (see
discussion in Appendix \ref{appendix_thermalization}). The constant
offset $y_{0}$ results from a slight inhomogeneity of the magnetic
field, which gives rise to a weak saddle potential. This increases
(decreases) the cloud size in y-direction (x-direction) during
expansion.
%**********************************************************************************
\section{Experimental Results}
\label{results}
%**********************************************************************************
Here we first discuss our measurements of the frequency $\omega_{\rm
q}$ and the damping rate $\kappa$ of the in-trap oscillation
(Sec.\,\ref{freq_damp}). We then present the data for the phase
offset $\phi$ and the amplitude $A$
(Sec.\,\ref{surface_subsection_phase_amp}). Finally, we explore the
hydrodynamic-to-collisionless transition
(Sec.\,\ref{surface_temperature_dependence}). As commonly used in
the field of BEC-BCS crossover physics \cite{Varenna2006}, the
dimensionless parameter $1/k_{\rm F}a$ is introduced to characterize
the interaction regime. The parameter $k_{\rm F} = \sqrt{2mE_{\rm
F}}/\hbar$ is the Fermi wave number and
$m$ is the mass of an atom.\\
%*************************************************************************************
\subsection{Frequency and damping}
\label{freq_damp}
%*************************************************************************************
\begin{figure}[!h]
\center{
\includegraphics[width=9cm]{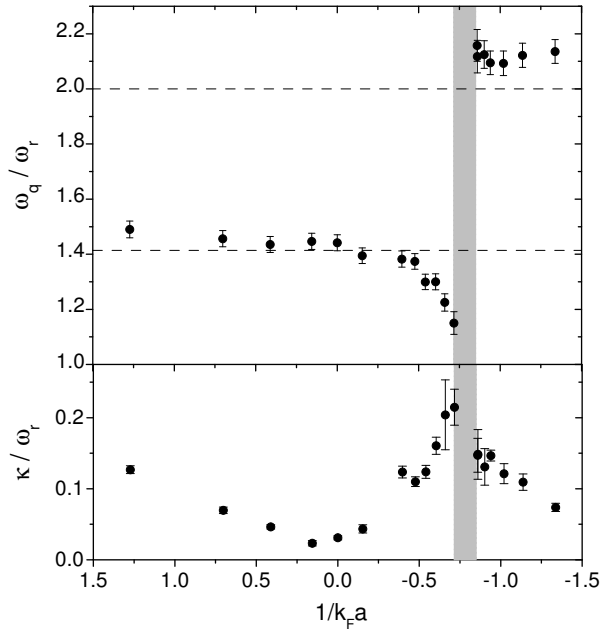}}
\caption{Frequency $\omega_{\rm q}$ (upper plot) and damping rate
$\kappa$ (lower plot) of the radial quadrupole mode. Both quantities
are normalized to the radial trap frequency $\omega_{\mathrm{r}}$
and plotted versus the interaction parameter $1/k_{\rm F}a$. The
dashed lines indicate the theoretical predictions in the
hydrodynamic ($\omega_{\rm q}/\omega_{\mathrm{r}} = \sqrt{2}$) and
in the collisionless limit ($\omega_{\rm q}/\omega_{\mathrm{r}} =
2$). The shaded area marks the transition from hydrodynamic to
collisionless behavior between $1/k_{\rm F}a \approx -0.72$
($B\approx 930$\,G) and $1/k_{\rm F}a \approx -0.85$ ($B\approx
960$\,G).} \label{surface_freq_damping}
\end{figure}
In Fig.\,\ref{surface_freq_damping}, we show the results for the
frequency $\omega_{\rm q}$ and the damping rate $\kappa$ of the
radial quadrupole mode throughout the BEC-BCS crossover. Both
$\omega_{\rm q}$ and $\kappa$ are normalized to the trap frequency
$\omega_{\rm r}$, which we determine by a sloshing mode measurement
\cite{Altmeyer2006a}. We include small corrections resulting from
anharmonicity of the trapping potential and the residual ellipticity
of the trap (see appendix \ref{appendix_corrections}).

The data confirm the expected transition between the hydrodynamic
and the collisionless regime on the BCS side of the resonance (see
Sec.\,\ref{theory}). The transition is qualitatively different from
the hydrodynamic-to-collisionless crossover in a classical gas
\cite{Buggle2005} or in a Fermi gas without superfluidity
\cite{Vichi2000a}. Instead of a continuous and monotonous variation
of the frequency between the two limits
($\sqrt{2}\,\omega_{\mathrm{r}}$ and $2\,\omega_{\mathrm{r}}$), an
abrupt change occurs. When this transition is approached from the
hydrodynamic side, a striking frequency downshift shows up as a
precursor of the transition to higher frequencies. In the transition
region (shaded area in Fig.\,\ref{surface_freq_damping}), no data
points are shown because of the large damping and correspondingly
very large uncertainties for the measured frequency.

 The damping rate shows similar behavior as in our previous measurements on the radial
compression mode \cite{Bartenstein2004b,Altmeyer2006b}. Maximum
damping occurs near the hydrodynamic-to-collisionless transition,
whereas minimum damping is observed slightly below the resonance. In
general, we find that damping is roughly two times larger for the
quadrupole mode than for the compression mode at the same
temperature \cite{endnote_damping}. The faster damping of the
quadrupole mode is plausible in view of the larger frequency change
at the transition.

We now discuss the behavior in different regions in more detail:
\begin{description}
\item[\it$1/k_{\rm F}a \approx 0$\,\rm:]In the unitarity
limit, the normalized frequency agrees well with the theoretically
expected value of $\omega_{\rm q}/\omega_{\mathrm{r}} = \sqrt{2}$
for a hydrodynamic gas, see
Eq.~\eqref{paper_surface_freq_hydrodynamic}. To check for
consistency with previous experiments \cite{Altmeyer2006a}, we here
also reproduced the frequency $\sqrt{10/3} \, \omega_{\mathrm{r}}$
of the radial compression mode on the $10^{-3}$ accuracy level. The
damping is low for the Fermi gas in the unitarity limit. In contrast
to the compression mode, the quadrupole mode frequency stays
constant throughout the crossover, indicating that it is independent
of the equation of state.

\item[\it$1/k_{\rm F}a > 0$\,\rm:]In the strongly interacting BEC
regime, there is an increase in the damping and a slight increase in
the frequency for increasing $1/k_{\rm F}a$. As the gas is more
susceptible to heating by inelastic processes in the deep molecular
regime \cite{Varenna_Rudi2006}, both effects may be due to a thermal
component in this region.

\item[\it$1/k_{\rm F}a \approx -0.8$\,\rm:]The frequency exhibits
the pronounced ``jump'' from the hydrodynamic to the collisionless
frequency. This transition is accompanied by a pronounced maximum of
the damping rate.

\item[\it$1/k_{\rm F}a \lesssim -0.8$\,\rm:]The frequency
stays almost constant about 5\% above the theoretically expected
value of $\omega_{\rm q} = 2\omega_{\mathrm{r}}$. Interaction
effects in the attractive Fermi gas are likely to cause this
significant upshift \cite{Pedri2003a,Urban2006}. As we cannot
experimentally realize a non-interacting Fermi gas above the
resonance, we could not perform further experimental checks.

\item[\it$1/k_{\rm F}a \lesssim0$ \rm{and} $1/k_{\rm F}a \gtrsim
-0.8$ \,\rm:] In this regime, we detect a substantial down shift in
the quadrupole mode frequency. The effect begins to show up already
slightly above the resonance  ($1/k_{\rm F}a = 0$) and increases to
a magnitude of almost 20\% ($\omega_{\rm q}/\omega_{\mathrm{r}}
\approx 1.15$ at $1/k_{\rm F}a = -0.72$), before the transition to
collisionless behavior occurs. Indications of a similar down shift
have been observed already in compression mode experiments
\cite{Bartenstein2004b,Altmeyer2006b,Kinast2004a}, but here the down
shift is considerably larger and not blurred by changes in the
equation of state.
\end{description}

A plausible explanation for the curious behavior of the collective
mode frequency on the BCS side of the resonance is provided by
coupling of the oscillation to the pairing gap
\cite{Combescot2004a,Chin2004a,Varenna_Rudi2006}. If we assume that
the abrupt transition is caused by pair breaking resulting from
resonant coupling of the oscillation to the gap, then the down shift
may be interpreted as a coupling effect
 when the gap is not much larger
than the oscillation frequency \cite{Combescot2006}. A similar shift
may also arise from coupling of hydrodynamics and quasiparticle
motion \cite{Urban2006}. The observed phenomenon still awaits a full
theoretical interpretation.

%*************************************************************************************
\subsection{Phase shift and amplitude}
\label{surface_subsection_phase_amp}
%*************************************************************************************
Additional information on the interaction regime is provided by the
phase shift $\phi$ and the amplitude $A$ of the observed oscillation
(see Eq.\eqref{surface_fit}). This is useful since extremely high
damping in the transition region makes a meaningful determination of
frequency and damping practically impossible. We find that both
amplitude and phase shift, however,  can be determined with
reasonable uncertainties even in the transition regime.\\
\begin{figure}[h]
\center{
\includegraphics[width=9cm]{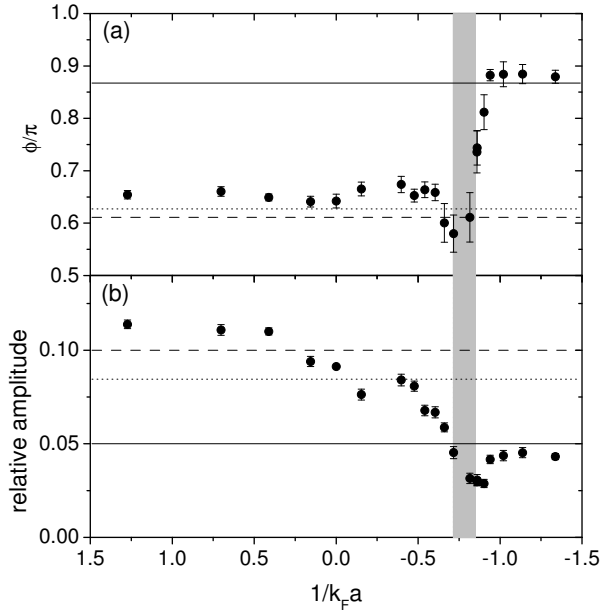}}
\caption{ (a) Phase shift $\phi$  and (b) relative amplitude  of the
quadrupole mode versus interaction parameter $1/k_{\rm F}a$ after
$t_{\mathrm{TOF}} = 2$ms expansion. The horizontal lines show
calculations from our theoretical model: the solid lines in the
collisionless limit, the dotted lines in the hydrodynamic regime at
unitarity ($\gamma = 2/3$) and the dashed lines in the hydrodynamic
regime in the BEC limit ($\gamma = 1$). These calculated values can
be read off from Fig.\,\ref{surface_expansion_measurements} for the
phase and Fig.\,\ref{surface_expansion_amplitude} for the amplitude.
The shaded area marks the transition between hydrodynamic and
collisionless behavior between $1/k_{\rm F}a \approx -0.72$ and
$1/k_{\rm F}a \approx -0.85$ (see also
Fig.\,\ref{surface_freq_damping}).} \label{surface_phase_amp}
\end{figure}
In the following, we present measurements of phase shift and
amplitude. These are compared to model calculations, which are
described in detail in appendix \ref{appendix_scaling}.

In Fig.~\ref{surface_phase_amp}, the phase $\phi$ and the relative
amplitude are plotted versus the interaction parameter $1/k_{\rm
F}a$. The relative amplitude is given by the amplitude $A$
(definition see Eq.\eqref{surface_fit}) divided by the average width
of the cloud after expansion. The average width is obtained by
averaging $(W_{\rm x}+ W_{\rm y})/2$ over one oscillation period
using the same data set from which we extract $A$.

In the transition area around $1/k_{\rm F}a = -0.8$, the phase shift
$\phi$ shows the step-like change at the transition from the
hydrodynamic to the collisionless regime. This is similar to the
jump in frequency in Fig.\,\ref{surface_freq_damping}. In the
collisionless and unitary regimes, the phase agrees with the
theoretically expected values (solid line and dotted line,
respectively).

As a general trend, the relative amplitude is larger in the
hydrodynamic and smaller in the collisionless regime. In the
hydrodynamic regime, the relative amplitude decreases for decreasing
$1/k_{\rm F}a$, which is explained by the change of $\gamma$ from 1
to 2/3; $\gamma$ is the polytropic index of the equation of state
(see Appendix\,\ref{appendix_scaling}). At unitarity, the relative
amplitude agrees well with the numerically calculated value for
$\gamma = 2/3$ (dotted line).  In the collisionless limit, the
relative amplitude is half of the value at unitarity, which is also
consistent with our calculations in App. \ref{appendix_amp_phase}.
We note that at the transition from the hydrodynamic to the
collisionless regime, the value of the relative amplitude decreases
even below the collisionless value.

In summary, the behavior of the phase shift and the amplitude agrees
with our model presented in Appendix \ref{appendix_scaling} (see
also Fig.\,\ref{surface_expansion_amplitude} and
Fig.\,\ref{surface_expansion_measurements}), in particular the
prominent change in the phase offset is confirmed.
%*************************************************************************************
\subsection{Further observations}
\label{surface_temperature_dependence}
%*************************************************************************************
The measurements presented in the preceding subsections were taken
under fixed experimental conditions, where only the scattering
length $a$ was varied. In this subsection we investigate how the
transition from hydrodynamic to collisionless behavior depends on
the experimental parameters excitation amplitude, trap depth and
temperature.

In a first set of experiments, we explored whether the position of
the transition depends on the excitation amplitude. We increase or
decrease the amplitude by a factor of 2. This allows us to compare
the oscillations where the amplitude is $\sim 20~$\%, $\sim 10~$\%
and $\sim 5~$\% of the averaged width. We do not observe any
significant change in the position of the transition.

In general, we find that the transition always occurs when the mode
frequency is similar to the pairing gap. This is supported by the
fact that when we vary the trap depth the transition occurs at a
constant scattering length ($a \approx -5000 a_{0}$, B $\approx
960$G) and does not depend on $1/k_{\rm F}a$
\cite{endnote_kfashift}. A change in laser power of our trapping
laser influences both Fermi energy $E_{F}$ and the frequency
$\omega_{\rm q}$. As we increase the trap power by a factor of 10,
we also increase the radial trap frequencies by a factor of
$\sqrt{10} \approx 3.2$. This changes the Fermi energy by a factor
of 2.2 and the pairing gap, which scales like the trap frequencies,
by roughly a factor of 3 \cite{Chin2004a}. These findings suggest
that the transition is linked to a coupling of the collective
oscillation to the pairing gap. This is also in agreement with
earlier results on the
radial compression mode \cite{Bartenstein2004b,Varenna_Rudi2006}.\\
\begin{figure}[h]
\center{
\includegraphics[width=9cm]{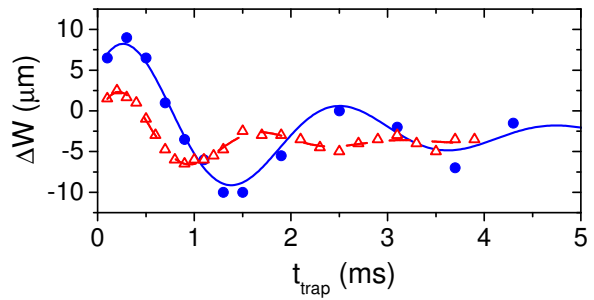}}
\caption{(color online) Oscillations of the quadrupole surface mode
at a magnetic field of $920$ G and $1/k_{\rm F}a = -0.66$. The
filled circles correspond to a cold ensemble, whereas the open
triangles correspond to a heated ensemble. The solid lines are fits
to the data according to Eq.\eqref{surface_fit}.}
\label{surface_temperature}
\end{figure}

To explore the temperature dependence of the transition between the
hydrodynamic and the collisionless phase, we use a controlled
heating scheme similar to the one described in \cite{Altmeyer2006a},
where we hold the gas in a recompressed trap and let it heat up. We
set the magnetic field to 920G ($1/k_{\rm F}a = -0.66$), i.e.
slightly below the hydrodynamic-to-collisionless transition, where
the regime is still clearly hydrodynamic. We observe the
oscillations in a gas at the lowest temperature we can achieve in
our experiments (filled circles) and in a ``hotter'' gas (open
triangles) in Fig.~\ref{surface_temperature}. The temperature of the
cold gas is $\lesssim 0.1\,T_{\rm F}$ and we believe the temperature
of the heated gas to be $\lesssim 0.2\,T_{\rm F}$.
Figure~\ref{surface_temperature} clearly shows that the frequency
for the colder ensemble is lower than that of the heated one and the
amplitude is lower by roughly a factor of 2. Using our model in
Appendix\,\ref{appendix_scaling} this indicates a temperature driven
transfer of the ensemble from the hydrodynamic to the collisionless
regime.

Thus we find that the radial quadrupole mode is suited to detect
temperature induced changes of the collisional regime of the gas. An
exploration of the phase diagram of our system depending on
temperature is possible, but beyond the scope of this article. In
our lab, work is  currently in progress on the radial scissors mode,
which turns out to be an even better tool for the exploration of
temperature effects.

%*******************************************************************
\section{Conclusions}
\label{conclusions}
%*******************************************************************
We have presented measurements on the radial quadrupole mode of an
ultracold $^{6}$Li Fermi gas in the BEC-BCS crossover. As a pure
surface excitation, this elementary mode probes hydrodynamic
behavior without being affected by changes in the equation of state.
We have measured the characteristic properties of this collective
mode in a wide range of interaction strengths.

Our observations provide new insight into the dynamics of the gas,
in particular on the BCS side of the crossover, where the character
of the oscillations abruptly changes from hydrodynamic to
collisionless behavior. Our measurements presented in this paper
show the phenomenon much clearer than in the radial compression mode
\cite{Bartenstein2004b,Kinast2004a,Altmeyer2006b} and provide
quantitative data on the behavior near the transition. In
particular, the data show that a substantial down shift of the
collective mode frequency occurs in the hydrodynamic regime as a
precursor of the transition.

The experimental results support the interpretation that the
coupling of oscillation mode and pairing gap
\cite{Combescot2004a,Chin2004a,Varenna_Rudi2006} plays a crucial
role for the collective excitation dynamics on the BCS side of the
crossover. We anticipate that our new quantitative data on the
hydrodynamic-to-collisionless transition will stimulate further
theoretical investigations on this intriguing phenomenon.

%**********************************************************************
\section*{Acknowledgements}
%**********************************************************************
We thank S.\ Stringari and M.\ Urban for stimulating discussions. We
thank E. R. S$\mathrm{\acute{a}}$nchez Guajardo for helpful
discussions during the process of writing this paper. We acknowledge
support by the Austrian Science Fund (FWF) within SFB 15 (project
part 21). S.R.\ is supported within the Doktorandenprogramm of the
Austrian Academy of Sciences. M.J.W. is supported by a Marie Curie
Incoming International Fellowship within the 6th European Community
Framework Program (40333).

%************************************************************************
\begin{appendix}
\section{Scaling approach and expansion effects}
\label{appendix_scaling}
%************************************************************************
Here we present a theoretical model to describe the oscillation of
the cloud in the trap as well as its expansion after release; the
model adopts the scaling approach applied in
\cite{Bruun2000b,Menotti2002,kinastphd}. The interplay between the
dynamics of the collective mode and the expansion behavior is of
particular interest as it introduces novel methods to investigate
the collisional regime. We use a scaling approach for both the
hydrodynamic and the collisionless regime
\cite{Bruun2000b,Menotti2002,kinastphd}. In
App.\,\ref{appendix_hydro}, the limit of a hydrodynamic gas is
presented, whereas in App.\,\ref{appendix_collisionless}, the model
in the collisionless regime is discussed. Based on these models, we
show calculated results for the amplitude and the phase after
expansion in App.\,\ref{appendix_amp_phase}.

The scaling approach describes the cloud at the time $t$ after
excitation \cite{Bruun2000b,Menotti2002,kinastphd}. Using the
scaling function $b_{i}(t)$ for $i = x,y$, the width $W_{i}(t)$ for
all times $t>0$ can be written as
\begin{equation}
\label{expansion_width} W_{i}(t) = b_{i}(t)W_{i}(0),
\end{equation}
where $W_{x}(0) = (1-\alpha)W_{0}$ and $W_{y}(0) = (1+\alpha)W_{0}$
are the initial widths at excitation and $W_{0}$ is the width of the
cloud without excitation. The initial conditions for the scaling
function are $b_{i}(0) = 1$ and
$\dot{b}_{i}(0) = 0$.\\
%********************************************************************************
\subsection{Dynamic behavior in the hydrodynamic limit}
\label{appendix_hydro}
%********************************************************************************
In the hydrodynamic limit, the equations of hydrodynamics lead to
the following differential equations for $b_{x}$ and $b_{y}$
\cite{Menotti2002}
\begin{eqnarray}
\label{expansion_hydro_difeq} \ddot{b}_{x} &=& \frac{\omega^{2}_{0
x}}{b_{x}\left(b_{x}b_{y}\right)^{\gamma}} -
b_{x}~\omega^{2}_{x},\nonumber \\
 \ddot{b}_{y} &=& \frac{\omega^{2}_{0
y}}{b_{y}\left(b_{x}b_{y}\right)^{\gamma}} - b_{y}~\omega^{2}_{y},
\end{eqnarray}
where $\gamma$ is the polytropic index of the equation of state and
$b_{z}(t) = 1$ for our elongated trap geometry. The parameters
$\omega_{0x}$ and $\omega_{0y}$ are the trap frequencies at the
moment of excitation ($t = 0$), when the cloud has no further
excitation and is in thermal equilibrium. In contrast to this,
$\omega_{\mathrm{x}}(t)$ and $\omega_{\mathrm{y}}(t)$ are the time
dependent trap frequencies. The timing scheme is illustrated in
Fig.\,\ref{surface_timing}. The following equation summarizes the
behavior of the trap frequencies $\omega_{i}(t)$:
\begin{equation}
\label{time_dependent_trap_freq} \omega_{i}(t) = \left\{
              \begin{array}{ll}
                   \omega_{0i} & , \,t = 0\\
                   \omega_{\mathrm{r}} & , \,0< t < t_{\rm trap}\\
                   0 & ,\, t > t_{\rm trap}.
              \end{array}
       \right.
\end{equation}
This enables us to calculate the scaling functions $b_{x}$ and
$b_{y}$ as solutions of Eq.\eqref{expansion_hydro_difeq} for the
in-trap oscillation. In the limit of small amplitudes ($\alpha \ll
1$) solutions are
\begin{eqnarray}
\label{solutions_hydro} b_{x} &=& 1 +\alpha(1- \cos{\omega_{\rm q} t}),\nonumber\\
 b_{y} &=& 1 -\alpha(1 - \cos{\omega_{\rm q} t}),
\end{eqnarray}
where $\omega_{\rm q} = \sqrt{2}\,\omega_{\mathrm{r}}$ is the radial
quadrupole oscillation frequency. Together with
Eq.\eqref{expansion_width}, we are able to determine the difference
in widths of the cloud to be
\begin{equation}
\label{delta_W_hydro} \Delta W = -2\,\alpha W_{0}\cos{\omega_{\rm q}
t}.
\end{equation}

%******************************************************************************************
\subsection{Dynamic behavior in the collisionless limit}
\label{appendix_collisionless}
%******************************************************************************************
In the collisionless limit, the following set of two uncoupled
equations characterizes $b_{i}$, where $i$ stands for $x,y$,
\cite{Bruun2000b}
\begin{equation}
\label{collisionless_diff_eq_limit} \ddot{b}_i =
\frac{\omega_{0i}^2}{b_{i}^3} - b_i\omega_i^2.
\end{equation}
In the limit of small amplitudes ($\alpha \ll 1$) solutions of the
in-trap oscillation are
\begin{eqnarray}
\label{solutions_coll} b_{x} &=& 1 +\frac{\alpha}{2}(1- \cos{\omega_{\rm q} t}),\nonumber\\
 b_{y} &=& 1 -\frac{\alpha}{2}(1 - \cos{\omega_{\rm q} t}),
\end{eqnarray}
where $\omega_{\rm q} = 2\,\omega_{\mathrm{r}}$ is the radial
quadrupole oscillation frequency. Together with
Eq.\eqref{expansion_width}, we are able to determine the difference
in widths of the cloud to be
\begin{equation}
\label{delta_W_coll} \Delta W = -\alpha
W_{0}\left(1+\cos{\omega_{\rm q} t}\right).
\end{equation}
In contrast to the hydrodynamic limit, the oscillation is initially
not centered around $\Delta W = 0$. Furthermore the oscillation has
an amplitude 1$/$2
of the amplitude in the hydrodynamic gas.\\
\begin{figure}[h]
\center{
\includegraphics[width=8cm]{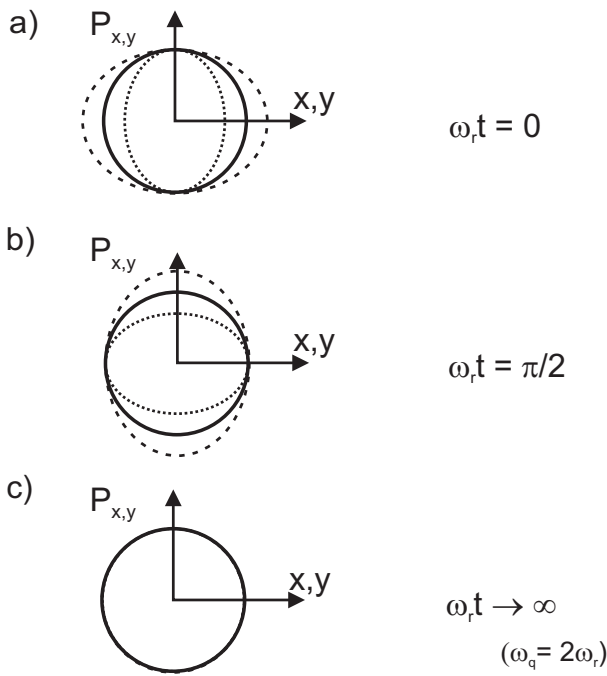}}
\caption{Phase space dynamics for the quadrupole mode in the
collisionless regime. Shown are phase space contours of an ensemble
of particles which is held in a round trap (i.e. $ \omega_x =
\omega_y = \omega_{\rm r} $). In (a) and (b) the situation during
the oscillation in the trap is shown for two different times $t$.
The solid line indicates the equilibrium phase space contour
(without excitation), whereas the dotted (dashed) line shows the
contour in the x (y) direction after excitation of the oscillation
mode. (c) After long times, residual thermalization finally damps
out the oscillations and leads to a circular phase space contour.}
\label{collisionless_phase}
\end{figure}
Besides the finding of analytical solutions, it is enlightning to
understand the collective oscillations in the collisionless limit by
considering the phase space dynamics of the cloud. In
Fig.~\ref{collisionless_phase}, we show the contours of phase space
distributions in the x- and y- directions. The axes are scaled such
that for the round trap, i.e. $ \omega_{\rm x} = \omega_{\rm y} =
\omega_{\rm r} $, the dynamics of any point in phase space is a
simple circular rotation about the origin with frequency
$\omega_{\rm r}$. Thus, the solid circle in
Fig.~\ref{collisionless_phase} (a) indicates an equilibrium phase
space contour for the round trap. Right after applying the
excitation scheme as described in Sec.\,\ref{experimental procedure}
the phase space contours in the x- and y- direction are given by the
dashed and dotted ellipses in Fig.~\ref{collisionless_phase} (a).
Since the gas is fully thermalized at the instant of excitation, the
initial momentum distribution in x- and y- direction is the same. As
time progresses, the elliptic contours will rotate with frequency
$\omega_{\rm r}$ (see Fig.~\ref{collisionless_phase} (b)), which
corresponds to oscillations in the trap. We note that both the
spatial and the momentum distribution in the x-direction are never
larger than the ones in the y-direction. Therefore, $\Delta W$
oscillates between $2 \alpha W_0$ and zero and the aspect ratio of
the cloud never inverts. This is to be compared to the hydrodynamic
case where $\Delta W$ oscillates between $\pm 2 \alpha W_0$.

Residual thermalization effects in a near collisionless gas will
damp out the initial oscillation amplitude of $ \alpha W_0$ and one
will eventually end up again with a circular phase space contour
(see Fig.~\ref{collisionless_phase} (c)). This is studied in detail
in Appendix\,\ref{appendix_thermalization}.

%*************************************************************************
\subsection{Amplitude and phase on expansion}
\label{appendix_amp_phase}
\begin{figure}[h]
\center{
\includegraphics[width=9cm]{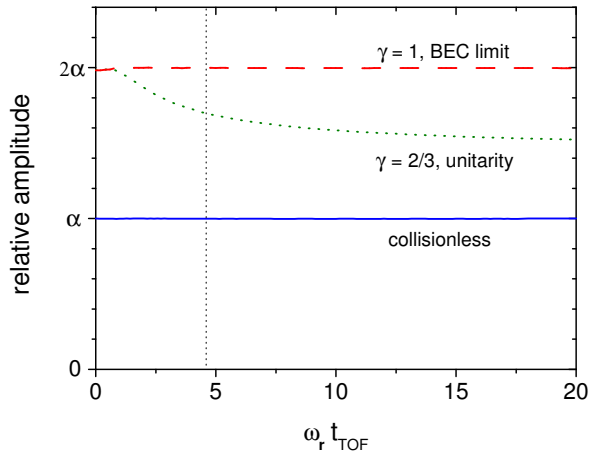}}
\caption{(color online). Calculated relative amplitude  of a surface
mode oscillation versus reduced time of flight $\omega_{\rm r}
t_{\rm TOF}$ after release from the trapping potential. The values
are calculated for the hydrodynamic (dashed curve: $\gamma=1$,
dotted curve: $\gamma=2/3$) and collisionless regime (solid curve).
The vertical dotted line marks the typical expansion time in our
experiments.} \label{surface_expansion_amplitude}
\end{figure}

\begin{figure}[h]
\center{
\includegraphics[width=9cm]{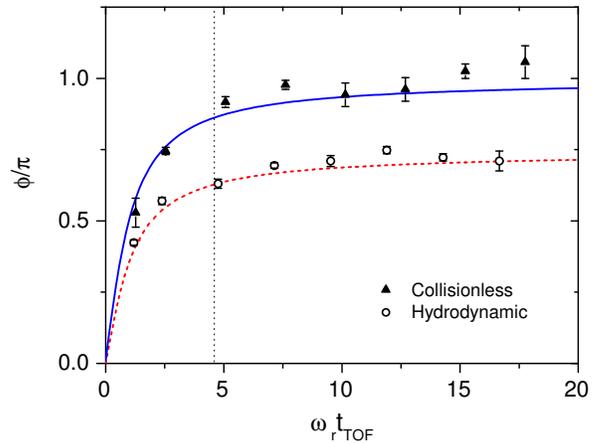}}
\caption{(color online) Phase $\phi$ of the collective surface mode
as detected by fits according to Eq.\eqref{surface_fit} versus
reduced expansion time $\omega_{\rm r}t_{\mathrm{TOF}}$ at unitarity
(open circles) and at $1/k_{\rm F}a = -1.34$ (filled triangles). The
lines are numerical simulations for the hydrodynamic (dashed line)
and collisionless regime (solid line). The vertical dotted line
marks the typical expansion time in our experiments.}
\label{surface_expansion_measurements}
\end{figure}

%*************************************************************************
Here we present our calculated results based on the models in
App.\,\ref{appendix_hydro} and App.\,\ref{appendix_collisionless}
for the hydrodynamic and the collisionless limit, respectively. We
show the relative amplitude that is given by the amplitude $A$
(definition see Eq.\eqref{surface_fit}) divided by the average width
of the cloud after expansion (for definition details see
Sec.\,\ref{surface_subsection_phase_amp}). Calculations of this
relative amplitude are shown in
Fig.\,\ref{surface_expansion_amplitude}, whereas calculations and
measurements for the phase offset $\phi$ are shown in
Fig.\,\ref{surface_expansion_measurements}.\\
Fig.~\ref{surface_expansion_amplitude} shows the calculated relative
amplitude of a surface mode oscillation in the hydrodynamic (dashed
and dotted curves) and in the collisionless (solid curve) regime as
function of the reduced expansion time $\omega_{\rm
r}t_{\mathrm{TOF}}$. The hydrodynamic curves are calculated for the
BEC limit of $\gamma = 1$ (upper, dashed curve) and in the unitarity
limit of $\gamma = 2/3$ (lower, dotted curve). The amplitude in the
collisionless regime is smaller than in the hydrodynamic regime.
Initially the amplitude of the excitation is half as large in the
collisionless as in the hydrodynamic regime, as already explained in
App.\,\ref{appendix_collisionless}. In expansion the normalized
amplitude stays constant in the collisionless regime and in the
hydrodynamic regime for $\gamma = 1$. For $\gamma = 2/3$ in
the hydrodynamic regime it decreases for longer expansion times.\\

In Fig.~\ref{surface_expansion_measurements} we compare experimental
data for the phase shift $\phi$ with numerical simulations. The data
have been taken at unitarity where $1/k_{\rm F}a = 0$ (hydrodynamic,
open circles), and on the BCS-side of the resonance at $1/k_{\rm F}a
= -1.34$ (collisionless, closed triangles). The dashed line is based
on a model for the hydrodynamic interaction regime and the solid
line on a model for the collisionless regime. The data agree with
the theoretical model where no free fit parameters are used. This
confirms our approach presented above.
%------------------------------------------------------------------------------------------
\section{Thermalization effects in a near-collisionless gas}
\label{appendix_thermalization}
%------------------------------------------------------------------------------------------
Here we describe thermalization effects in a near-collisionless gas
that are not included in the model for the collisionless limit in
App.\,\ref{appendix_collisionless}. Despite the word
``collisionless'', collisions play a crucial role for thermalization
for our experimental parameters. A typical time scale for
thermalization processes is only a few oscillation cycles long. By
analyzing the theory, we are able to introduce a universal fit
function, as given by Eq.\eqref{surface_fit}, which describes the
oscillation both in the hydrodynamic and in the near-collisionless
regime.

The measured behavior of the nearly collisionless quadrupole
oscillation (see Fig.\,\ref{surface_typical_oscillation}) has two
characteristics: after excitation the oscillation is centered around
$\Delta W = (W_{\rm x}(0)-W_{\rm y}(0))/2$, then after some time it
is centered around $\Delta W = 0$. These two limits are consistent
with thermalization of the gas on a relevant time scale greater than
the period of the oscillation.

In order to model these effects, we follow a theory based on a
classical gas in the transition between the hydrodynamic and the
collisionless behavior described in \cite{Pedri2003a}. An
application of this theory for the compression mode in the
hydrodynamic regime has been used in \cite{kinastphd}. Here we will
handle thermalization effects of the quadrupole mode in the
near-collisionless regime.

Using the classical Boltzmann-Vlasov kinetic equation in the
relaxation-time approximation and ignoring mean field effects one
can derive  the following coupled differential equations
\cite{Pedri2003a}
\begin{equation}
\label{collisionless_diff_eq_b} \ddot{b}_i =
\omega_{0i}^2\frac{\theta_{i}}{b_i} - \omega_i^2b_i
\end{equation}
and
\begin{equation}
\label{collisionless_diff_eq_theta} \dot{\theta_i}
 = \frac{1}{\tau_R}(\theta_i-\bar{\theta})-
2\frac{\dot{b_i}}{b_i}\theta_i .
\end{equation}
The parameter $b_i$ is the scaling function described earlier in
Appendix \ref{appendix_scaling}; $\theta_i$ is a scaling parameter
directly related to the temperature and $\bar{\theta} = \frac{1}{3}
\Sigma_k \theta_k$. The initial condition for $\theta_i$ is
$\theta_i(0) = 1$, as long as the gas is in thermal equilibrium at
the moment of the excitation. The parameter $\tau_R$ is the
relaxation time which describes the time scale of collisions. In the
collisionless limit, when $\tau_R \rightarrow \infty$, the
differential equations \eqref{collisionless_diff_eq_b} simplify to
the simple form in Eq.\,\eqref{collisionless_diff_eq_limit}. For the
hydrodynamic limit ($\tau_R \rightarrow 0$), we find
Eq.\,\eqref{expansion_hydro_difeq}.\\

\begin{figure}[h]
\center{
\includegraphics[width=9cm]{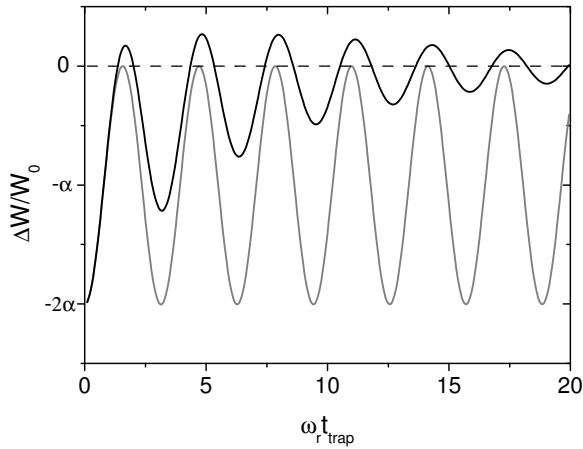}}
\caption{Calculated quadrupole oscillations in the
near-collisionless regime.  The lines show the relative difference
in width $\Delta W$ as a function of the reduced time $\omega_{\rm
r}t_{\rm trap}$. The oscillation is modeled according to
Eq.\eqref{collisionless_diff_eq_b} and
\eqref{collisionless_diff_eq_theta}. The dark line shows the result
of the calculation when $\omega_{\rm r}\tau_R  = 2.3$ and the grey
line shows the coscillation in the collisionless limit at
$\omega_{\rm r}\tau_R = 1000$.}
\label{collisionless_oscillation_theory}
\end{figure}
The solutions to these equations depend on the parameter $\tau_R$ as
can be seen in Fig.\,\ref{collisionless_oscillation_theory}. Our
measured data in the collisionless regime are well described by
$\omega_{\rm r}\tau_R
 \sim 2.3$ (compare to
Fig.~\ref{surface_typical_oscillation}).

%********************************************************************************
\subsubsection*{The universal fit function}
\label{appendix_fitfunction}
%********************************************************************************

We find that the model calculations from
\eqref{collisionless_diff_eq_b} and
\eqref{collisionless_diff_eq_theta} can be well described with the
following fit function
\begin{eqnarray}
\label{appendix_surface_fit}
\begin{split}
 \Delta W = ~&
A~e^{-\kappa t_{\rm trap}}~\cos{(\omega_{\rm q}t_{\rm trap}
+\phi)}\\& + C~e^{-\xi t_{\rm trap}}+y_{0}.
\end{split}
\end{eqnarray}
The first term describes the exponentially damped oscillations. The
second term describes the shift of the center of the oscillation in
the collisionless regime. The third term $y_{0}$ is a constant
offset which will be discussed later.

We have used Eq.\eqref{appendix_surface_fit} to fit our experimental
measurements.  We find that the free fit parameters $\xi$ and
$\kappa$ are related through $\xi/\kappa \approx 1.5$ for all our
measurements in the near-collisionless regime. In the hydrodynamic
regime $C=0$, and therefore $\xi$ becomes irrelevant.

The constant offset $y_{0}$ is due to an experimental artifact
 that results from a slight inhomogeneity of the magnetic field.
At the location of the atoms the inhomogeneous magnetic field leads
to a weak saddle potential which causes a slight anisotropic
expansion during time of flight. This anisotropy is responsible for
a slight offset in $\Delta W$.\\

%--------------------------------------------------------------------
\section{Corrections to the normalized frequency}
\label{appendix_corrections}
%--------------------------------------------------------------------
The theoretical normalized frequencies $\omega_{\rm q}/\omega_{\rm
r}$ are calculated for perfectly harmonic trapping potentials in an
idealized symmetric trap geometry. There are small derivations from
this conditions in real experiments. In order to compare the
experimental data to the idealized theoretical case, we have to
correct our data. The measured normalized frequency $\omega_{\rm
q}/\omega_{\rm r}$ of the radial quadrupole mode has to be increased
because of two small corrections. The larger correction is based
upon a slight anharmonicity of the trapping potential and the
spatial extension of the cloud in the trap. The smaller correction
is caused by a small residual ellipticity of the trapping potential.

The potential created by our trapping beam has a Gaussian shape.
This results in a nearly harmonic potential in the center of the
trap; however, for higher precision one must take into account
higher order terms of the potential. Anharmonicity effects influence
both our measurements of the sloshing mode frequency, where we
determine $\omega_{\rm r}$, and our measurements of the quadrupole
mode frequency $\omega_{\rm q}$. As we evaluate the normalized
frequency $\omega_{\rm q}/\omega_{\rm r}$, the anharmonicity effects
on sloshing and quadrupole mode almost cancel out each other. The
small remaining correction to the normalized frequency is included
by multiplying with a prefactor $1+ b\sigma$
\cite{Altmeyer2006a,dr_alex}. The anharmonicity parameter $\sigma$
relates the energy of the oscillation to the total potential depth
and is defined by $\sigma = \frac{1}{2}m \omega_{r}^2 r_{\rm rms}^2
/V_0$, where $r_{\rm rms}$ is the root-mean-square radius of the
trapped cloud and $V_0$ is the potential depth. The parameter $b$
depends on the interaction regime. In the hydrodynamic regime, it is
given by $(4+ 10\gamma)/(2+7\gamma)$, whereas in the collisionless
regime $b$ is determined by $6/5$ \cite{dr_alex}. Here, $\gamma$ is
the polytropic index of the equation of state. In our experiments,
typically $b\sigma \approx 0.014$ , but $b\sigma$ can rise to an
upper limit of $b\sigma < 0.027$.

In the hydrodynamic regime, there is also a correction due to
residual ellipticity effects. This correction takes into account
that we compare our measurements with a theory for non-elliptic
geometries. The ellipticity $\epsilon$ of the trap is defined by
$\epsilon = (\omega_{\rm y} - \omega_{\rm x})/\omega_{\rm r}$. In
our experiments, the ellipticity is small and given by $\epsilon
\approx 0.07$. Therefore, we can apply the ellipticity correction by
multiplication of a prefactor $1+\lambda\epsilon^{2}$
\cite{Altmeyer2006a,dr_alex}, where the interaction dependent factor
$\lambda$ is given by $(\gamma + 2)/(4\gamma)$. Altogether,
$\lambda\epsilon^{2}$ is smaller than 0.006 for all data points.
\end{appendix}
%*******************************************************************


\begin{thebibliography}{39}
\bibitem[{mod()}]{modulationnote}
\bibinfo{note}{We use a periodic modulation with an arc sine-like function. This results in much better harmonic potentials than a simple sawtooth modulation ramp \cite{diplom_christoph}.}
\end{thebibliography}
\end{document}